\documentstyle[prl,twocolumn,aps]{revtex}
\begin{document}
\preprint{}
\title{Avalanches and Correlations in Driven Interface Depinning
}
\author{Heiko Leschhorn}
\address{
Theoretische Physik III,
Ruhr-Universit\"at Bochum,
Universit\"atsstr. 150, D-44801 Bochum, Germany
}
\author{Lei-Han Tang}
\address{
Institut f\"ur Theoretische Physik, Universit\"at zu K\"oln,
Z\"ulpicher Str. 77, D-50937 K\"oln, Germany
}

\date{\today}
\maketitle
\begin{abstract}
We study the critical behavior of a driven interface in a medium
with random pinning forces by analyzing spatial and
temporal correlations in a lattice model recently proposed
by Sneppen [Phys. Rev. Lett. {\bf 69}, 3539 (1992)].
The static and dynamic behavior of the model is related to
the properties of directed percolation.
We show that, due to the interplay of local and global
growth rules, the usual method of dynamical scaling has to
be modified. We separate the local from the global part of
the dynamics by defining a train of causal growth events, or
"avalanche", which can be ascribed
a well-defined dynamical exponent $z_{loc} = 1 + \zeta_c
\simeq 1.63$ where $\zeta_c$ is the roughness exponent of the
interface.

\end{abstract}
\narrowtext
\section{Introduction}

The behavior of a driven interface subjected to quenched random
forces
plays an important role in the ordering kinetics of impure magnets
and other domain growth phenomena \cite {review1}.
The driving force $F$ can be realized by a magnetic field,
pressure or chemical potential favoring the growth of one of the
coexisting phases. If $F$ is weak, the interface is typically pinned
in one of the many locally stable configurations. In this case growth
is possible only through thermally activated hopping,
which is an extremely slow process at low temperatures.
However, when $F$ exceeds some critical value $F_c$,
all metastable states disappear and the interface is then free to
move even at zero temperature.
The depinning of the interface at $F_c$ can be considered as a
critical
phenomenon where characteristic quantities show power-law behavior,
e.g. the velocity
of the interface is expected to scale as $v \sim (F-F_c)^\theta $,
where $\theta $ is a critical exponent.
It is known that charge-density waves pinned by impurities exhibit
very similar behavior \cite {cdwfish}.
The dynamical behavior associated with the
disappearance of metastable states
(i.e. avalanche) as $F$ is increased
towards $F_c$ has been a subject of recent interest
in the study
of systems far from equilibrium
\cite {soc,earthquake,rob,cdwmidd}.

A plausible continuum description of the
interface dynamics is given by the following equation with
a Kardar-Parisi-Zhang (KPZ) \cite {kpz} nonlinear term,
$$ {\partial h \over \partial t} =  \nu
\nabla ^2 h + {\lambda \over 2} (\nabla h)^2 + F
- \eta ({\bf x} , h), \eqno(1) $$
where $h({\bf x},t)$ is the height of the interface.
Unlike the original KPZ-equation for Eden-type growth processes,
here $ \eta ({\bf x},h)$ is a quenched random force with short range
correlations.
In the case $\lambda = 0 $, all critical exponents
describing the depinning transition have been
calculated recently in a functional renormalization
group treatment close to
four interface dimensions \cite {nstl}.
For growth in an isotropic medium, it is plausible that the
$\lambda$-term is not present when the interface moves with
vanishing velocity. However, the term can be present for anisotropic
growth. Most solid-on-solid type models are expected to be in the
latter category.

In 1+1 dimensions, there is a particularly simple class of lattice
models
which exhibit a critical depinning transition \cite{tl,boston}.
The mechanism for
pinning is the directed percolation of cells with pinning forces
$\eta$ greater than the driving force $F$. The threshold force $F_c$
needed to depin the interface is then simply related to the
critical percolation density $\rho _c$ of such cells by
$F_c = 1 - \rho _c$.
The roughness exponent of the pinned interface is equal to that
of the critical percolation cluster,
$\zeta = \zeta _c \simeq 0.63 $.

In a separate development, Sneppen introduced a simple model
(model B of Ref. \cite{snep}, hereafter referred to as the Sneppen
model)
to examine the interplay between local and global rules of growth
in determining interface roughening and temporal correlations.
He found numerically in (1+1)-dimensions, that the roughness of
the interface also obeys the scaling of a string on a critical
directed percolation cluster.

As we explain below, the same roughness exponent for the two types
of models is due to the fact that the
same geometrical constraint (the ``Kim-Kosterlitz'' condition)
is invoked in defining a locally stable configuration \cite {tlcom}.
There are however differences in the way the interface is driven.
Although the pinning and elastic forces on a given site
in the Sneppen model are {\it local}
as in eq.(1),
each time the site on the interface with minimal
$\eta=\eta_{\rm min}$ is selected {\it globally }
and made to grow by one unit in height.
Neighboring interface sites then adjust themselves to recover the
geometrical constraint. In the language of a uniformly driven
interface,
such a rule corresponds to increasing $F$ just above $\eta_{\rm min}$
to make one site unstable, and then quickly set $F$ to a much
smaller value to prevent an avalanche taking place.

The Sneppen rule allows one to sample a particular sequence of
interface configurations, each of them being a metastable
configuration
at some value of $F$. Successive configurations in the sequence
differ
only by an infinitesimal amount, i.e. a few sites (about 4) which
have moved when one site is made unstable. The situation here
resembles that of an interface at a finite temperature and driven by
a uniform force $F$ far below $F_c$: Local irreversible motions are
made possible by thermal fluctuations, but the interface stays always
close to some metastable state.

Further numerical studies of the model by Sneppen and Jensen
\cite {sj}
revealed interesting spatial correlation of successive growth events.
In addition, they have found that, in the saturated regime,
the height advance
at a given column exhibits complicated scaling with time, with
exponents varying with the moment considered.

The aim of the present paper is to analyze the spatial and temporal
correlations in the Sneppen model and try to relate the observed
scaling behavior to the properties of directed percolation.
It turns out that some of these correlations depend on the value
of $\eta _{\rm min}$ (or equivalently $F$) at a given moment.
Since $ \eta _{\rm min}$ fluctuates in time, for these correlations
the temporal translational invariance is lost. This is an important
feature of the dynamics based on global rules.

Another significant consequence of the global rules is that
growth is no longer homogeneous in space. At a given moment,
only a small part of the interface is moving. The growth events
that follow may either closeby, or far away. Due to this property,
the usual method to perform dynamical scaling, based on the
presumption
of a single growing correlation length, should be modified.
In this connection
we found it useful to
distinguish growth events which are closeby and hence bear strong
correlations, from those which are far apart.
We observe that a
train of growths events, started with some $\eta _{\rm min}^0$,
propagates laterally with a
well-defined dynamical exponent $z_{loc}$, which can be related to
the roughness of the directed percolation cluster:
$z_{loc} = 1 + \zeta _c \simeq 1.63$.
In the context of a driven interface,
this motion can be thought of as an avalanche.
We found that the distribution of avalanche sizes obeys a
power-law decay up to a size related to $\eta _{\rm min}^0$.
The spatial-temporal correlations
between successive growth events, on the other hand include both
local and global motion.

The paper is organized as follows. In section II we recall the
definition of the Sneppen model and present
a theorem which relates the stable (static)
configurations of the Sneppen model to
directed percolating strings.
In Section III we define the avalanches (causal events) and determine
their
dynamical behavior as well as their size distribution. In Section IV
the spatial-temporal correlations are investigated by considering the
distribution of lateral distances between successive growth events.
Section V contains conclusions and a summary.

\section {Distribution of pinning forces and roughness}

We first review the definition of the Sneppen model \cite {snep}.
Each cell $(i,h)$ on a square lattice is assigned
a random pinning force $\eta (i,h)$ uniformly distributed in the
interval
$[0,1)$. The interface is specified by a set of integer column heights
$h_i ~(i=1,...,L)$ with the
local slope constraint $|h_i-h_{i-1}|\leq 1$ for all
$h_i$ ("Kim-Kosterlitz" condition).
Growth $h_j\rightarrow h_j+1$ proceeds at the site $j$
where the pinning force $\eta(j, h_j)=\eta_{\rm min}$ is the minimum
among
all interface sites, followed by necessary adjustments at neighboring
sites until the slope constraint is recovered.
The growth rules are illustrated in Fig.1.

Since we want to relate the behavior of the interface in the
Sneppen model to directed percolation, we first
recall some of the properties of directed percolation clusters
\cite {percgen}.
When the density $\rho$ of occupied sites is less than some threshold
value $\rho _c$, a typical cluster of occupied sites connected
horizontally or diagonally extends over a distance
of the order of $\xi_\parallel$ in the parallel direction
and a distance of the order of $\xi_\perp$ in the perpendicular
direction. For $\rho >\rho_c$, there appears a directed percolating
cluster which extends over the whole system. This cluster has
a network structure of nodes and compartments, where
each compartment has an anisotropic shape similar
to the connected clusters below $\rho _c$, characterized by
$\xi_\parallel$ and $\xi_\perp$.
On both sides of the percolation transition, the two
lengths have the power-law behavior
$$\xi_\parallel\sim |\rho-\rho_c|^{-\nu_\parallel},\qquad
\xi_\perp\sim |\rho-\rho_c|^{-\nu_\perp}.\eqno(2)$$
Series calculations give
$\nu_\parallel=1.733\pm 0.001~$, $\nu_\perp=1.097\pm 0.001$
\cite {percexp}
and $\rho_c = 0.5387 \pm 0.003 $ \cite {percfc}.
The roughness of a percolating string scales as
$\xi_\perp \sim \xi_\parallel ^{\nu_\perp / \nu_\parallel}$,
i.e. the roughness exponent $\zeta _c = \nu_\perp / \nu_\parallel
\simeq 0.63.$

In Ref.\cite {tlcom} we proposed to study
the distribution $P_p (\eta) $ of pinning forces
$\eta (i,h_i) $ at the interface and the probability distribution
$P_m (\eta_{\rm min})$ during growth.
When starting at $t=0$ with a flat interface $h_i \equiv 0$, the
forces $\eta (i,h_i) $
are equally distributed. During the transient regime the interface
becomes rough and since
always the smallest pinning force $\eta_{\rm min}$ is updated,
the sites with small $\eta (i,h_i) $ get rare.
This in turn implies that the typical value of the selected
$\eta _{\rm min}$
increases with time.
The distributions $P_p (\eta) $ and
$P_m (\eta_{\rm min})$
shown in Fig.2a were recorded in the transient regime
in the time interval
$L/4 \leq t < L/2$ for a system of size $L=8192$.
The peak of $P_m (\eta_{\rm min})$ moves to the right with increasing
time and thereby "eating up the store" of small
$\eta (i,h_i) $
which were present at $t=0$ for the flat interface.
As long as there were no $\eta_{\rm min}$ larger than
a value $\eta _u (t) $,
the distribution $P_p (\eta)$ is still a constant for
$\eta > \eta _u (t) $.
(In Fig.2a $\eta _u (t) \approx 0.3$.)
In the next paragraph we show that in the thermodynamic limit,
there will never be an
$\eta _{\rm min} $ larger than a critical value
$ F_c = 1 - \rho_c \simeq 0.461$
and therefore the transient regime ends when
$\eta _{\rm min}$
first comes close to $F_c$.
When the peak of
$P_m (\eta_{\rm min})$
approaches $F_c$, its height vanishes and the distribution becomes
stationary, which we show in Fig.2b together with $P_p(\eta)$ in the
saturated regime. Since $\eta _{\rm min} \leq F_c$, the stationary
distribution $P_p ( \eta )$ is flat for $\eta > \eta_u = F_c$
in the limit $L \to \infty$.

To see that the growing interface always has
$\eta_{\rm min} \leq F_c$ in the thermodynamic limit we first note
that every interface configuration satisfying the slope constraint
is a path on a directed percolation cluster
of sites with $\eta$ greater than or equal to $\eta_{\rm min}$.
Such a path only exists if the density $1-\eta_{\rm min}$ of these
cells on the lattice is greater than the critical
percolation density $\rho_c$, i.e.
$\eta_{\rm min}\leq 1-\rho_c $ for all interfaces.
Paths on the infinite {\it critical} percolating cluster
have the {\it largest} $\eta _{\rm min} = F_c = 1 - \rho_c
\simeq 0.461 $.

In a numerical simulation with a finite system however,
we see only a part of an infinite critical percolating cluster.
Thus, an interface
which traces out a critical path can have a value $\eta_{\rm min}$
slightly larger than $F_c$ and the distribution
$P_m (\eta_{\rm min})$ in the saturated regime in Fig.2b is not
exactly zero for $\eta _{\rm min} > F_c$.
The motion of the flat interface at
$t=0$ to the first
critical path corresponds to the
transient regime (see Fig.2a).

In the following we show that an
interface in the transient regime as well as
an interface which already crossed a critical path, is driven
to configurations with successively increasing $\eta _{\rm min}$
and thereby approaching the next path on a critical percolating
cluster.
We first introduce a few notations.
Assuming that the random forces $\eta$ are real numbers,
there will not be two interface configurations sharing the same
$\eta _{\rm min}$
because an interface is always updated at the site with
$\eta _{\rm min}$.
Thus each interface configuration
$\{ h_i \} $ can be characterized by its
$\eta _{\rm min}$
and is denoted by
$ H[\eta _{\rm min}]$.
An order relation is defined by
$ H[\eta _{\rm min}^A] > H[\eta _{\rm min}^B]$
if $h_i^A \geq h_i^B$ for all $i$ and
$ H[\eta _{\rm min}^A] \neq H[\eta _{\rm min}^B]$.
Consider an interface
$ H[\eta _{\rm min}^0]$
at a time $t=t_0$ and choose a real number $c$ with
$\eta _{\rm min}^0 \leq c \leq F_c$.

We next show that the growing interface at times $t > t_0$
has to overlap completely with the closest percolating path
which has $\eta _{\rm min} > c$.
This closest path is
defined by
$ H[\eta _{\rm min}^c] \equiv
{\rm min} \bigl \{ ~H[\eta _{\rm min}] $, such that
$\eta _{\rm min} > c$ and
$ H[\eta _{\rm min}] > H[\eta _{\rm min}^0] ~\bigr \}$.
There can be many percolating pathes which share the same
$\eta _{\rm min}^c$ but only the lowest path
$ H[\eta _{\rm min}^c] $
will be realized by the interface as can be seen as follows.

The growing interface configurations
$ H[\eta _{\rm min}] $ at times $t \geq t_0$ which are below
$ H[\eta _{\rm min}^c] $,
$ H[\eta _{\rm min}^0] \leq H[\eta _{\rm min}] <
H[\eta _{\rm min}^c] $, all have an
$\eta _{\rm min}<\eta _{\rm min}^c$
by the definition of $H[\eta _{\rm min}^c] $.
(This is also true for interfaces
$ H[\eta _{\rm min}] $
which already overlap in part with
$ H[\eta _{\rm min}^c] $, no matter if
the site with $\eta _{\rm min}^c $ is already on the interface
$ H[\eta _{\rm min}] $.)
Therefore these $\eta_{\rm min}$ will all be updated
{\it before} $\eta _{\rm min}^c$
and hence the interface will grow until
$ H[\eta _{\rm min}] = H[\eta _{\rm min}^c]$ and then update the
site with $\eta _{\rm min}^c$ because
any interface site can advance at most by
one unit in height upon each updating.

The motion with $c=\eta_{\rm min}^0$ we will call later an "avalanche"
(see Sec.III).
If we choose however $c=F_c$,
it follows that all pathes which are the minimum on a
{\it critical } percolation cluster, i.e.
$ H[\eta _{\rm min}^c] = H[F_c]$ in the thermodynamic limit, act
as ``checking points'' where the interface has to go through.
There, a "snapshot" of the distribution $P_p (\eta)$ is exactly a
step function. It can be seen from Fig.2b
that the intermediate
interface configurations between two checking points
are also close to critical, in the sense that only a small
percentage (which we expect to vanish in the thermodynamic limit)
of sites on the interface have $\eta<  F_c$.
Hence the interface in the Sneppen model has approximately the
roughness exponent $\zeta$  of the
critical directed percolating path with
$\zeta \simeq \zeta _c \simeq 0.63$.

We have seen that the
interface $ H[\eta _{\rm min}]$ is successively driven
to configurations $ H[\eta _{\rm min}^c]$ with
$\eta _{\rm min}^c > \eta _{\rm min}$
and thereby eliminating $\eta (i,h_i) < \eta _{\rm min } ^c $
on the interface. This process (see Fig.2)
may be perceived as "self-organizing" part of the approach
to criticality, where $\eta_{\rm min}^c = F_c$ for infinite
systems.
In comparison, the interface in the model of Ref. \cite {tl}
is pinned by the critical percolating cluster when the driving
force $F$ is tuned to its threshold value $F_c = 1 - \rho _c$.

The roughness exponent $ \zeta $ can be measured by
the equal time height-height correlation function
$C(r) = \langle \overline{[h(r+r',t) - h(r',t)]^2} \rangle
\sim r^{2 \zeta} $
where the overbar and the angular brackets denote the spatial and
the configurational average, respectively.
For a system of size
$L=8192$
we found a roughness exponent
$\zeta = 0.655 \pm 0.005$ which is somewhat larger than the
critical value 0.63. However, the measured exponent $\zeta$
varies systematically with
the system size: For $L=900$ we get $\zeta = 0.665 \pm 0.005$,
whereas for $L=65536$ we found $\zeta = 0.648 \pm 0.005 $.
An explanation is that in a finite
system the distribution $P_p (\eta)$ is not exactly a
step function and the motion between two critical clusters
yields an effective exponent larger than 0.63 similar to
a moving interface of Ref. \cite {tl}.
In our simulations,
$ P_p (\eta) $ approaches the step function with increasing
system size.
Thus we expect that the measured discrepancy to the
exponent of the percolating cluster
$\zeta _c \simeq 0.63$ vanishes in the thermodynamic limit.
This is also supported by a simulation where we measured $C(r)$
only when $\eta_{\rm min} $ is close to $F_c$. In this case
$\zeta = 0.64 \pm 0.01$ which is consistent with the expected
critical value.

\section {Causal events: avalanches}

After a transient regime, when $\eta _{\rm min}$
first comes close to $F_c$,
the interface in the Sneppen model exhibits
a steady-state critical behavior (saturated regime),
which allows a convenient study
of the dynamics at criticality. However, the behavior is complicated,
caused by the interplay between the local adjustments due to the slope
constraint and the rule that the growth site with
$\eta = \eta _{\rm min}$ is chosen among all interface sites.
Successive growth events can be far apart and the motion is therefore
inhomogeneous in space. Hence, a single growing
correlation length does not exist
and the usual dynamical scaling has to be considered with care.
The realized sequence of growth sites after a time $t=t_0$
depends on the globally chosen value $\eta_{\rm min} (t=t_0)$,
which is responsible for the growth inhomogeneity in space.
Thus we will separate the local
part from the global part of the dynamics by defining an "avalanche",
which has the property that the sequence of growth sites {\it inside}
the avalanche does {\it not} depend on
$\eta_{\rm min} (t=t_0)$.

An avalanche is defined by a sequence of growth events (including
the necessary adjustments due to the slope constraint)
started at any integer time $t=t_0$ where
$ \eta_{\rm min} (t=t_0) $ is denoted by
$ \eta_{\rm min} ^0 $.
This avalanche is terminated at the first time $\tau$
when the successive $\eta_{\rm min} (\tau +1) $ is larger than
$ \eta_{\rm min} ^0 $, i.e.
for all times $t$ with $t_0 < t \leq \tau $ the growth events have
$ \eta_{\rm min} (t) <  \eta_{\rm min} ^0 $.
We call this
causal events because the avalanche consists of a train of
growths events which are all
induced by local adjustments in $h_i$ due to the slope constraint
after $t=t_0$.

To see in what sense the sequence of growth sites inside an avalanche
is independent of $ \eta_{\rm min} ^0 $, we consider
at $t=t_0$ two identical interface configurations $A$ and $B$ with
the same random forces $\eta (i,h) $  above the interface but
with different $\eta (i,h_i)$ at the interface such that
$ \eta_{\rm min} ^0 [A] <  \eta_{\rm min} ^0 [B] $
at the same site $j$.
For the configuration $A$ there exist forces $\eta (i,h_i)$ with
$ \eta_{\rm min} ^0 [A] < \eta (i,h_i) <  \eta_{\rm min} ^0 [B] $
but for interface $B$ all $\eta (i,h_i) > \eta_{\rm min} ^0 [B] $.
Since the random forces above the interface are assumed to be the
same, all $ \eta_{\rm min} $ of the growing interfaces $A$ and $B$
inside both avalanches are identical because they are
all induced by identical local adjustments after
$t=t_0$.
When however at time $\tau ^A +1$,
$ \eta_{\rm min} ^0 [A] < \eta_{\rm min} (\tau ^A +1)
<  \eta_{\rm min} ^0 [B] $,
i.e. when avalanche $A$ is terminated, this new growth site of
interface $A$ can be far away, but for the interface $B$ it
has to be still induced by the local adjustments
of avalanche $B$ because there were no
$ \eta (i,h_i) < \eta_{\rm min} ^0 [B] $ at $t=t_0$.
We have seen that although the random environment at (and
below) the two interfaces
$A$ and $B$ is different, the motion inside the avalanche is
only influenced by the local adjustments after $t= t_0$.

A size $s$ of the avalanche can be defined by the number of growth
events including the necessary adjustments,
$s= \sum_i \bigl( h_i(\tau ) - h_i (t_0) \bigl )$,
and a width by
${\rm max} \{~i$, such that $h_i (\tau) > h_i (t_0)~\} ~-~
{\rm min} \{~i $, such that $h_i (\tau) > h_i (t_0)~\}$.

{}From Sec.II we know that an interface with
$\eta _{\rm min}^0$
is driven to configurations with
$\eta _{\rm min}^c > \eta _{\rm min}^0$.
At $t=t_0$ with $ \eta_{\rm min} (t=t_0) = \eta_{\rm min} ^0 $,
a part of the interface starts to move which was "pinned"
just before $t_0$ by a path with
$\eta (i,h_i) \geq \eta _{\rm min}^0$. This part of the interface
moves through a compartment of the percolation cluster
to the next path which again pins the interface with
$\eta (i,h_i) > c = \eta _{\rm min}^0$ (see Sec.II).
Since the compartment has a height of the order of
$\xi_\perp ( \eta_{\rm min} ^0) $ and a width of the order of
$\xi _\parallel ( \eta_{\rm min} ^0)$, it is a natural
conjecture that the width of the avalanche scales with
$ \xi _\parallel ( \eta_{\rm min} ^0) \sim
[F_c - \eta_{\rm min} ^0]^{-\nu _\parallel} $
and the size is at most
$\xi _\parallel ( \eta_{\rm min} ^0) * \xi _\perp (\eta_{\rm min} ^0)
\sim [F_c - \eta_{\rm min} ^0]^{-(\nu _\perp + \nu_\parallel)} $.

Note that at every (integer) time an avalanche is started and big
avalanches
can contain smaller ones. Thus successive growth events inside a
big avalanche can be quite far apart (jumps between small avalanches),
but all events are inside the correlation length
$ \xi_\parallel (\eta_{\rm min} ^0)
\sim [F_c - \eta_{\rm min} ^0]^{-\nu _\parallel} $.
In this sense an avalanche is "localized" and we call the
corresponding motion "local dynamics".
The presumption of dynamical scaling,
that there is only a single growing correlation length, is
reestablished for the local dynamics and we
can ascribe a well-defined local dynamical exponent $z_{loc}$ to
the lateral propagation of growth inside the avalanche:
$\tau  \sim \xi _\parallel ^{z_{loc}}$. Since the size of an
avalanche is proportional to the time $\tau$, one has from
$\xi_\parallel ^{z_{loc}} \sim \tau \sim \xi_\perp \xi_\parallel $
$$ z_{loc} \simeq 1+ \nu_\perp / \nu_\parallel =
1 + \zeta _c \simeq 1.63. \eqno (3) $$

In a simulation, this local dynamical exponent can be detected by
considering the infinite moment of the height-height time correlation
function in the saturated regime
$$ C_q (t) = \left \langle \overline {[ h_i (t+t') - h_i (t') -
\overline {h_i (t+t') - h_i (t') } ]^q } ^{1/q} \right \rangle ,
\eqno (4) $$
which becomes for $q \to \infty $
$$ C_\infty (t) \simeq \left \langle {\rm max}_i\{h_i (t+t') -
h_i (t')\} \right  \rangle . \eqno (5)  $$
Since only the column with
maximum height-advance $\Delta h_{\rm max}$
contributes to the infinite moment
$C_\infty (t)$,
most growth events during the time $t$ are involved
in the motion through a compartment of a percolation cluster
of height $\Delta h_{\rm max}$, i.e.
$\Delta h_{\rm max} \sim \xi_\perp \sim t/\xi_\parallel
\sim t^{1-1/z_{loc}}=t^{\zeta / z_{loc}}$.
Thus
$C_\infty (t) \sim t ^ {\beta _ \infty}$  with
$\beta _\infty = \zeta / z_{loc} = \beta _{loc}$.
Sneppen and Jensen observed a scaling of
$C_\infty (t) $  with
$\beta _\infty = 0.41 \pm 0.02 $ which is close to the
exponent $\beta_{loc} = \zeta / z_{loc} =
\nu_\perp /(\nu_\perp + \nu_\parallel ) \simeq 0.39 $.
Our own simulations give $\beta _\infty = 0.40 \pm 0.01 $
which is in perfect agreement with $\beta_{loc}$ if we insert
the measured $\zeta \simeq 0.655 $.

The distribution $P_h(\Delta h,t)$ of height advances
$ \Delta h(t) = h_i (t'+t) - h_i(t') $ is shown in Fig.3a for
$2^6 \leq t \leq 2^{15}$ and $\Delta h>0$. $P_h$ has a large
peak at $\Delta h=0$ which is not shown, i.e. most of the columns
$i$ have not
grown (over 99 percent for $t=2^6$ and 70 percent for $t=2^{15}$ with
$L=8192$). Next we show that the distribution $P_h (\Delta h,t)$ for
the moving columns can be roughly brought to a "local" scaling form
when $\Delta h$ is scaled by $t^{\beta_{loc}}$. To normalize
$P_h(\Delta h>0,t)$
we note that the portion $n_{mov}$ of columns which have moved,
scales as the correlation length divided by the system size,
$n_{mov} \sim t^{1/z_{loc}} / L$. Thus one has
$$ P_h (\Delta h>0,t) = {t^{1/z_{loc}} \over L}~ {1 \over
t^{\beta_{loc}}}~
\Gamma \left ( {\Delta h \over t^{\beta_{loc}}} \right ) $$
$$ \sim t^{1-2\beta _{loc}}~
\Gamma \left ( {\Delta h \over t^{\beta_{loc}}} \right ), \eqno (6)$$
where $\Gamma (y)$ is a scaling function (see Fig.3b).
For fast growing columns with
large $\Delta h/t^{\beta_{loc}}$ the above argument for
$\Delta h _{\rm max} $ applies
and the data collapse in Fig.3b is perfect.
For smaller $\Delta h/t^{\beta_{loc}}$, however, there are
significant deviations from the "local" scaling form.

For the second moment $C_2 (t) $ \cite {difdef}
Sneppen and Jensen observed a scaling with an
exponent $\beta _2 = 0.69 \pm 0.02$ \cite {sj}.
However, due to the inhomogeneity in growth,
the application of dynamical scaling is questionable
as mentioned above.
The deviation of the effective exponent $\beta _2$ from the
scaling of the local dynamics is caused by the fact that
only a small part of the interface
($n_{mov} \sim t^{1/z_{loc}} / L$) has moved for short times $t$.
The observed value for
$\beta_2$ can be explained by
using that $\Delta h$ scales {\it roughly} with $t^{\beta_{loc}}$
for $\Delta h>0$ (see Fig.3b). For general integer $q$ we write
$$C_q(t) = \left \langle \overline {\left ( \Delta h - \overline
{\Delta h} \right ) ^q } ^ {1/q} \right \rangle $$
$$ = \left \langle \left ( \overline  {\Delta h ^q}
-q \overline {\Delta h^{q-1} } ~ \overline {\Delta h}
+~...~+ \overline {\Delta h}^q \right ) ^{1/q}
 \right \rangle $$
$$ \simeq
\left \langle \left ( {t^{1/z_{loc}} \over L} t^{q\beta_{loc}} -
q {t^{1/z_{loc}} \over L} t^{(q-1)\beta_{loc}}{t\over L}+...+
{t^q \over L^q} \right ) ^ {1\over q} \right \rangle  $$
($\overline {\Delta h (t)}  = t / L$).
In the observed scaling regime $t \ll L^{z_{loc}}$ and therefore
the main contribution comes from the first term.
Thus we have
$$ C_q (t) \sim t^{\beta_{loc} + 1/qz_{loc}} \sim
t^{1+(1-q)/ qz_{loc}}, \eqno (7)$$
i.e. $\beta _q \simeq 1+ (1-q)/q(1+\zeta)$ and $\beta_2
\simeq 0.70 $ which agrees with the simulations, while
for $q \to \infty ~$, $ \beta _q \to \beta_{loc}$.

Next the avalanche size is investigated.
We observe that the distribution
$ P_{av} ( s, \eta _{\rm min} ^0) $
of the avalanche size $s$ for a given
$\eta _{\rm min} ^0 $
shows a power-law decay with an exponent
$\kappa \simeq 1.25 \pm 0.05 $ up to a size
$s_0 \sim
\xi _\parallel ( \eta_{\rm min} ^0) * \xi _\perp (\eta_{\rm min} ^0)$
and then drops to zero for $s> s_0$ (see Fig.4a).
Thus the avalanche size distribution obeys the scaling form
$$ P_{av}(s) = s ^{- \kappa} ~ \Phi \left ( {s \over  s_0}  \right )
\eqno (8)$$
with $\Phi (y) = const.$ for $y<1$ and a rapid decay for $y>1$.
The good data collapse onto the scaling form eq.(8)
in Fig.4b supports our picture that
an avalanche corresponds to the motion through a compartment
of a percolation cluster, which is characterized by the lengths
$\xi_\perp $ and $\xi_\parallel$.

\section {Spatial-temporal correlations}

In this section we try to understand the
spatial-temporal correlations between successive growth events. To
this end we investigate the probability distribution
$P_{co} (x, \Delta t) $,
where $x$ is the distance parallel to the interface
between growth events which occur after a time $\Delta t$.
Sneppen and Jensen \cite {sj} observed that
$P_{co} (x, \Delta t) $ is constant for sufficiently small $x$
and a has power-law decay with an exponent $\gamma = 2.25 \pm 0.05 $
above a value $x_c$ which increases with $\Delta t$ (Fig.5a).

We next explain that the behavior
$P_{co} (x, \Delta t)  = const $ corresponds to the dynamics
of causal growth events inside an avalanche.
The local adjustments due to the slope constraint
after the avalanche has started induce randomly distributed
$ \eta (i,h_i) $. During the avalanche all
$\eta _{\rm min} (t) < \eta _{\rm min} ^0 $
are taken from these newly appeared $\eta (i,h_i) $.
Thus, the $\eta _{\rm min} (t) $ are randomly distributed
in space, i.e. the distance between successive growth events
is also equally distributed as long as $\Delta t$ is smaller than
the duration of the avalanche $\tau$, i.e. as long as
$ x < \xi_\parallel (\eta_{\rm min} ^0) $.
Therefore we can cast $P_{co} (x,\Delta t) $ into the scaling form
$$ P_{co} = { 1 \over \Delta t^{1/z_{loc}}} ~
\Psi \left ({ x \over  \Delta t^{1/z_{loc}}} \right ) \eqno (9) $$
with $z_{loc}= 1 + \zeta$
and $\Psi (y) = const. $ for $y<1$ and $\Psi \sim y^{-\gamma} $ for
$y>1$.
The scaling form eq.(9) with a satisfactorily data collapse is
shown in Fig.5b.

We see that the spatial-temporal correlations depend on the
value $\eta _{\rm min} $. For $\eta _{\rm min} $ close to
$F_c$, $x$ is equally distributed even for large $x$.
For small $\eta _{\rm min} $ on the other hand,
$P_{co} (x,\Delta t) $ obeys a power-law decay also for small $x$,
i.e. it is more probable that successive growth events are closeby.
Thus, for these correlations the temporal translational invariance is
destroyed.

The value of the exponent $\gamma $
can be understood by considering the conditional
distribution $P_{co}$ for fixed
$\eta _{\rm min}$
of the first event, which we denote by
$ \tilde P_{co} (\eta _{\rm min},x,\Delta t)$.
We express $ P_{co} (x,\Delta t)$ by
$$ P_{co} (x,\Delta t) = \int d \eta _{\rm min}~
 \tilde P_{co} (\eta _{\rm min},x,\Delta t)~
 P_{\eta}(\eta _{\rm min}).
\eqno (10)$$  From
our simulation (see Fig.2b) we see that
the probability distribution
$ P_{\eta}(\eta _{\rm min}) \sim (F_c - \eta_{\rm min}) $
close to $F_c$. From the above discussion we know that
$  \tilde P_{co} (\eta _{\rm min},x,\Delta t) \simeq 1/\xi_\parallel$
if $\xi_\parallel > x$.  Thus the integral eq.(10) takes the form
$$ P_{co} (x,\Delta t) = \int _{\xi_\parallel > x} d \eta _{\rm min}~
{1 \over \xi_\parallel}~
(F_c - \eta_{\rm min}) $$
from which one obtains
$P_{co} (x,\Delta t) \sim x ^{- \gamma}$ with
$$\gamma = 1 + (2 / \nu_\parallel)
\simeq 2.16. \eqno (11) $$
This is quite close to our numerical result
$\gamma = 2.20 \pm 0.05$.
We have also directly measured the distribution
$ \tilde P_{co} (\eta _{\rm min},x,\Delta t) $
to check our assumptions.
We found that
$ \tilde P_{co} (\eta _{\rm min},x,\Delta t = 1) $ first decreases
for small $x$ due to a high probability for choosing the next
$\eta _{\rm min}$ from the newly appeared $\eta$ from the local
adjustments. However, to explain the exponent $\gamma$ we are
interested in large $x$ (and thus in large $\xi_\parallel$),
for which we indeed observe a constant
$ \tilde P_{co} (\eta _{\rm min},x,\Delta t) $
for $x < \xi_\parallel (\eta _{\rm min})$.

\section {Conclusions and summary}

We have analyzed a number of spatial and temporal correlations in the
Sneppen model \cite {snep,sj}. The imposed
slope constraint $|h_i - h_{i-1}| \leq 1$ is the reason
for the relation of the static and dynamic behavior
to the properties of directed percolation.
The roughness exponent of the interface in the Sneppen model and
of the pinned interface in the model of Ref. \cite {tl}
is equal to that of a percolating string, $\zeta _c \simeq 0.63$.

The difference between the two models is, however, that in the
model of Ref. \cite {tl} the interface is driven by a uniform force
whereas in the Sneppen model there is a
self-tuned driving force which keeps the interface at the onset
of steady-state motion and therefore the interface shows critical
behavior. This is achieved by the rule, that
the site grows, which has the weekest pinning force $\eta_{\rm min}$
among all sites of the interface. This induces a nonlocal part in
dynamics. As a consequence, the motion of the interface
is inhomogeneous in space and the methods of dynamical
scaling are not applicable in a direct way
because there is no single growing correlation length.
Thus we have separated the local from the global part of the motion
by introducing an "avalanche", and assigned a
well-defined dynamical exponent
$z_{loc} = 1 + \zeta_c$ to the lateral propagation of the growth
inside an avalanche. We found that the size distribution of the
avalanches started with $\eta_{\rm min} ^0$
has a power-law decay with an exponent $\kappa \simeq 1.25$
up to a size $\xi_\parallel (\eta_{\rm min} ^0) *
\xi_\perp (\eta_{\rm min} ^0) $.

The spatial-temporal correlations where investigated by the
probability distribution $P_{co}(x,\Delta t)$ which shows a crossover
from a behavior determined by causal growth events
($P_{co}(x,\Delta t) = const.$) to a power-law decay with an exponent
$\gamma \simeq 2.2$
which can be also related to exponents of directed percolation.
We have seen that for the distribution $P_{co}$ the temporal
translational invariance is lost, which is due to the global part
of the dynamics.

Upon completion of the paper we became aware of an independent
work by Z. Olami, I. Procaccia, and R. Zeitak where ideas similar
to ours have been developed.

\section*{Acknowledgements} We thank Y. C. Zhang for a useful
discussion
and C. K\"ulske for helpful conservations and
critical remarks on the manuscript.
The work is supported in part by the Deutsche Forschungsgemeinschaft
under SFB 166 and 341.

\section*{Figure captions}

\begin{figure} \caption{
An interface configuration (filled circles and full lines)
is updated (big circles) at the site which has the smallest pinning
force $\eta _{\rm min}$. Four possible growth events
are shown including local adjustments (small circles and dashed
lines) due to the slope constraint $|h_i - h_{i-1}| \leq 1$.
Time is measured in units of such growth events.
Each interface configuration may be considered as a directed
percolating path.}
\end {figure}

\begin{figure} \caption{
Distribution $P_p (\eta) $ of all pinning forces
$\eta (i,h_i)$ at the interface (full line) and
probability distribution $P_m (\eta _{\rm min})$
(dashed line) for a system of size $L=8192$.
(a) The distributions in the transient regime are averaged over the
time interval $L/4 \leq t < L/2$ and over 1000 independent runs.
(b) $P_p (\eta) $ and $P_m (\eta _{\rm min})$
averaged over time in the saturated regime.
}
\end {figure}

\begin{figure} \caption {(a)
Distribution $P_h (\Delta h,t)$ of height advances $\Delta h>0$
for time differences $2^6 \leq t \leq 2^{15} $ $(L=8192)$.
Higher curves correspond to larger $t$.
(The same plotting symbol is used for data at a given time $t$.)
In addition, there
is a large peak of $P_h$ at $\Delta h=0$ which is not
shown. The curves are normalized if the peak at $\Delta h=0$
is included.
(b)  Scaling plot eq.(6) with the local exponent
$\beta_{loc} \simeq 0.40$.
}
\end {figure}

\begin{figure} \caption { (a)
Logarithmically binned distribution of
avalanche sizes $s$, where $P_{av} (s,\eta_{\rm min})$
is the density of events in the range $[s,2s)$.
Lower curves correspond to
smaller $\eta_{\rm min}$.
($0.30 \leq \eta_{\rm min} < 0.46$, $L=8192$).
$P_{av} (s,\eta_{\rm min})$ has a power-law decay with an exponent
$\kappa \simeq 1.25 \pm 0.05 $ up to a size
$s_0 \sim
\xi _\parallel ( \eta_{\rm min} ^0) * \xi _\perp (\eta_{\rm min} ^0)$
and then drops to zero for $s> s_0$.
(b) Scaling plot of the avalanche size distribution (eq.8).
}
\end {figure}

\begin{figure} \caption{(a)
Logarithmically binned distribution of distances $x$ between
two growth events, where
$P_{co} (x,\Delta t) $ is the density of $x$ in the range $[x,2x)$.
The two growth events occur after a time $\Delta t$
($1 \leq \Delta t \leq 256$, $L=8192$).
Curves with a larger $\Delta t$ have a wider plateau.
The power-law decay has an exponent $\gamma = 2.20 \pm 0.05$.
(b)
Scaling plot of the distribution
$P_{co}(x,\Delta t)$ eq.(9), where $x$ is scaled by
$\Delta t^{1/(1+\zeta)}$
using the measured value $\zeta = 0.655$.
}
\end {figure}


\begin{thebibliography}{99}

\bibitem{review1} {\it Dynamics of Ordering Processes in Condensed
Matter}, edited by S. Komura and H. Furukawa (Plenum, New York,
1988);
{\it Phase transitions and Critical Phenomena}, edited by C. Domb
and J. L. Lebowitz, Vol.8 (Academic, New York, 1983).

\bibitem {cdwfish} D. S. Fisher, Phys. Rev. Lett. {\bf 50}, 1486
(1983);
O. Narayan and D. S. Fisher, Phys. Rev. Lett. {\bf 68}, 3615 (1992).

\bibitem {soc} P. Bak, C. Tang, and K. Wiesenfeld,
Phys. Rev. Lett. {\bf 59}, 381 (1987).

\bibitem {earthquake} J. M. Carlson and J. S. Langer,
Phys. Rev. Lett. {\bf 62}, 2632 (1989);
J. S. Langer and C. Tang, Phys. Rev. Lett. {\bf 67}, 1043 (1991).

\bibitem {rob} N. Martys, M. O. Robbins, and M. Cieplak, Phys. Rev B.
{\bf 44}, 12294 (1991); H. Ji and M. O. Robbins, Phys. Rev. B
{\bf 46}, 14519 (1992).

\bibitem {cdwmidd} D. S. Fisher and A. A. Middleton, Phys. Rev. B
{\bf 47}, 3530 (1993); O. Narayan and A. A. Middleton,
to be published.

\bibitem{kpz} M. Kardar, G. Parisi, and Y.-C. Zhang,
Phys. Rev. Lett. {\bf 56}, 889 (1986).

\bibitem{nstl} T. Nattermann, S. Stepanow, L.-H. Tang, and
H. Leschhorn, J. Phys. II France {\bf 2}, 1483 (1992);
O. Narayan and D. S. Fisher, to be published.
For numerical results on eq.(1) with $\lambda = 0$ in lower
dimensions see H. Leschhorn, Physica A {\bf 195}, 324 (1993).

\bibitem{tl} L.-H. Tang and H. Leschhorn, Phys. Rev. A {\bf 45},
R8309 (1992).

\bibitem{boston} S. V. Buldyrev, A.-L. Barab\'asi, F. Caserta,
S. Havlin,
H. E. Stanley, and T. Vicsek, Phys. Rev. A {\bf 45}, R8313 (1992).

\bibitem{snep} K. Sneppen, Phys. Rev. Lett. {\bf 69}, 3539 (1992).

\bibitem{tlcom} L.-H. Tang and H. Leschhorn,
Phys. Rev. Lett. {\bf 70}, 3832 (1992).

\bibitem{sj} K. Sneppen and M.H. Jensen, Phys. Rev. Lett. {\bf 71},
101 (1993).

\bibitem{percgen} D. Stauffer and A. Aharony,
{\it Introduction to Percolation Theory}, 2nd edition,
(Taylor \& Francis, London, 1992);
W. Kinzel, in {\it Percolation Structures and Processes},
edited by G. Deutscher, R. Zallen, and J. Adler (A. Hilger, Bristol,
1983), p.425; S. Redner, Phys. Rev. B {\bf 25}, 5646 (1982);
J. Kert\'esz and D. E. Wolf, Phys. Rev. Lett. {\bf 62}, 2571 (1989).

\bibitem{percexp}
J. W. Essam, A. J. Guttmann, and K. De'Bell, J. Phys. {\bf A21}, 3815
(1988);
However, more recent transfer matrix calculations by
D. ben-Avraham, R. Bidaux, and L. S. Schulman
[Phys. Rev. A {\bf 43}, 7093 (1991)] gave
$\nu_\perp/\nu_\parallel=0.630\pm 0.001$.

\bibitem {percfc} J. A. M. S. Duarte, Physica A {\bf 189}, 43 (1992).
Note that the directed percolation clusters in the model of
Buldyrev et.al.
Ref. \cite {boston} have a somewhat different geometry yielding a
different
value $\rho_c$ but the same critical exponents $\nu_\perp$ and
$\nu_\parallel$.

\bibitem {difdef} The definition of $C_q(t)$ in
Ref. \cite {sj} is slightly different from that
we used in eq.(4) which yields the expression in eq.(5) in
the limit $q \to \infty $.
Since in eq.(4) we first take the power $1/q$ and then average
over time, we still average in eq.(5) over time
(or over the disorder).

\end {thebibliography}
\end {document}